\documentclass[aip,jap,reprint]{revtex4-1}
\usepackage[dvips]{color,graphics,epsfig,rotating}
\usepackage{graphicx}

\draft 

\begin{document}

\title{Crystal structure and electronic structure of quaternary semiconductors Cu$_2$ZnTiSe$_4$ and Cu$_2$ZnTiS$_4$ for solar cell absorber } 

\author{Xiaofeng Wang}
\author{Junjie Li}
\author{Zhenjie Zhao}
\author{Sumei Huang}\thanks{Electronic mail: smhuang@phy.ecnu.edu.cn}
\author{Wenhui Xie}\thanks{Electronic mail: whxie@phy.ecnu.edu.cn}

\affiliation{Engineering Research Center for Nanophotonics and Advanced Instrument, Department of Physics, East China Normal University, Shanghai, 200062, China}

\date{\today}

\begin{abstract}
 We design two new I2-II-IV-VI4 quaternary semiconductors Cu$_2$ZnTiSe$_4$ and Cu$_2$ZnTiS$_4$, and systematically study the crystal and electronic structure by employing first-principles electronic structure calculations. Among the considered crystal structures, it is confirmed that the band gaps of Cu$_2$ZnTiSe$_4$ and Cu$_2$ZnTiS$_4$ originate from the full occupied Cu 3$d$ valence band and unoccupied Ti 3$d$ conducting band, and kesterite structure should be the ground state. Furthermore, our calculations indicate that Cu$_2$ZnTiSe$_4$ and Cu$_2$ZnTiS$_4$ have comparable band gaps with Cu$_2$ZnTSe$_4$ and Cu$_2$ZnTS$_4$, but almost twice larger absorption coefficient $\alpha(\omega)$. Thus, the materials are expected to be candidate materials for solar cell absorber.
\end{abstract}


\maketitle 

\section{INTRODUCTION}
Main focus of current research in solar energy and thermoelectric power generation is to discover and develop efficient materials. An ideal thin-film solar cell absorber material should have a direct band gap around 1.3$\sim$1.5 eV with abundant, inexpensive, and nontoxic elements. Cu-base selenide and sulfide are considered to be a promising class of materials for solar-cell applications\cite{solar1,czts1,czts2,czts3,czts4}. Cu(In,Ga)Se$_2$ (CIGS) is one of the most promising thin-film solar cell materials, demonstrating an efficiency about 20\%\cite{solar1}. However, In and Ga are expensive components, and the band gap is usually not optimal for high efficiency CIGS solar cells. Currently, designing and synthesizing novel, high-efficiency, and low cost solar cell absorbers to replace CIGS has attracted much attention. Among the materials that have been investigated, the group I2-II-IV-VI4 quaternary semiconductors Cu$_2$ZnSnS$_4$ (CZTS) and Cu$_2$ZnSnSe$_4$ (CZTSe) compounds have drawn significant interest because they contain abundant and nontoxic elements Cu, Zn, Sn, S, and the band gap of CZTS is about 1.5 eV\cite{czsse-gaphse06,czsse-gapldau}, which is ideal for solar cell application.


Motivated by the same thought, in this letter we report two new I2-II-IV-VI4 quaternary semiconductors Cu$_2$ZnTiSe$_4$ (CZTiSe) and Cu$_2$ZnTiS$_4$ (CZTiS), in which Sn of CZTSe and CZTS is replaced by Ti. Ti is group IVB element and cheaper than Sn. Although its dioxide TiO$_2$ has been used in dye-sensitized solar cell\cite{tio2}, it has not been drawn attention to the series of group I2-II-IV-VI4 quaternary semiconductors yet. By employing first-principles electronic structure calculations, we show that the fundamental band gaps of CZTiSe and CZTiS are around 1.0$\sim$1.4 eV, and the direct band gaps at $\Gamma$ are round 1.2$\sim$1.6 eV, which are comparable with CZTSe and CZTS, thus CZTiSe and CZTiS might be useful materials for solar cell absorber.

\section{COMPUTATIONAL DETAILS}
We have systematically investigated the structural and electronic properties of CZTiSe and CZTiS by using a high precise first-principles all-electron full-potential linearized augmented plane wave (FLAPW) method based on the density-functional theory (DFT)\cite{wien2k}. Generalized gradient approximation (GGA) of Perdew-Burke-Ernzerhof (PBE)\cite{pbe96} is used in our calculations to obtain structural properties. The lattice constants $a$ and $c$ are optimized, as shown in Table I. Meanwhile, the atomic positions are relaxed until the force on each atom is smaller than 1mRy/a.u. The self-consistent calculations are considered to be converged only when the integrated charge difference between input and output charge density is less than 0.0001. We use at least 900 $k$-points for Brillouin-zone integrations. As the GGA underestimates the band gaps, we have performed DFT+U calculations by using Engel and Vosko GGA plus an onsite Coulomb interaction $U_d$ (EVU) to generate accurate band gap. Such an EVU method can correctly describe the band gap of $d$-electron zinc-blende semiconductors CuInSe2, ZnO, CZTS and CZTSe\cite{EVUapp,czsse-gapldau}. In all EVU calculations, the electronic correlation effects is considered with U$_{eff}$=4 eV for Cu, Zn and Ti, respectively. Furthermore, It should be mentioned that the EVU calculations have a little influence on structural properties.

\begin{figure}
\includegraphics[width=8cm,angle=0]{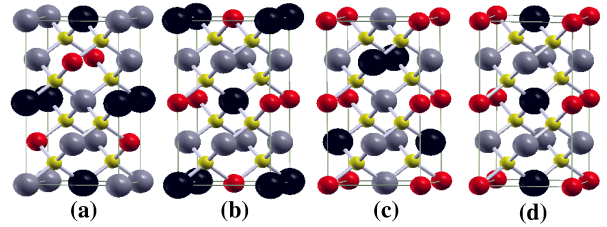}
\caption{\label{sum} (Color online) The crystal structures of CZTiSe (CZTiS) in (a) KS, (b) ST, (c) KS-I and (d) ST-I. Small yellow spheres are Se or S; middle red ones are Ti; big grey ones are Cu; largest black ones are Zn.}
\end{figure}

\section{RESULTS AND DISCUSSION}
The I2-II-IV-VI4 quaternary semiconductors have two fundamental crystal structures: one is kesterite structure (KS) [space group I\={4}, Fig.1(a)], the other is stannite structure (ST) [space group I\={4}2m, Fig. 1(b)]. We further consider two structures derived from KS and ST in which the position of Ti and Zn are exchanged every two planes, named as KS-I [space group P\={4}, fig. 1(c)] and ST-I [space group P\={4}2m, fig. 1(d)]. In all of structures, Se or S (group VI) atom is surrounded by two Cu (group I) atoms, one Ti (group IVB) atom and one Zn (group II) atom, therefore the octet rule is obeyed.

The calculated properties of CZTiSe and CZTiS of the four structures are listed in Table I. We find that KS compounds have the lowest total energy for both CZTiSe and CZTiS, whereas ST ones have lower total energy. Other two structures KS-I and ST-I should be unfavorable because of higher total energy. It seems that Ti and Zn are more likely to be tetrahedral coordination but not square coordination. Although the KS possesses a lower energy, the energy difference between KS and ST is small, about 2 meV/atom. It is similar with CZTSe and CZTS that KS is more stable than ST only with a tiny energy difference. Thus, it indicates that KS and ST ordering may coexist in real material, and the disorder effects should be similar with CZTS and CZTSe as discussed in Ref. 6. 

\begin{table}
\caption{Calculated lattice constant $a$, tetragonal distortion parameter $\eta$=$c/2a$, GGA band gap E$_{g}^{(GGA)}$, EVU band gap E$_{g}^{EVU}$ and the energy difference per atom $\delta E$ relative to the lowest-energy structure of different chalcogenides. }\label{struct}
\begin{tabular}{lccccccc}
\hline
\hline
 Sample  & Lattice &    $a$ (\AA)    &  $\eta$  & $\delta$$E$(meV)  & E$_{g}^{GGA}$(eV) & E$_{g}^{EVU}$(eV) &  \\
\hline
CZTiSe   &   KS      &   5.674    &  1.005     &   0   &  0.56  & 1.24   &\\
CZTiSe   &   ST      &   5.696    &  1.015     &  2.1  &  0.38  & 1.06   &\\
CZTiSe   &   KS-I    &   5.653    &  0.994     & 34.4  &  0.50  & 1.17   &\\
CZTiSe   &   ST-I    &   5.684    &  1.010     & 17.3  &  0.31  & 0.97   &\\

CZTiS    &   KS      &   5.381    &  0.998     & 0     & 0.68   & 1.42  &\\
CZTiS    &   ST      &   5.418    &  1.018     & 1.8   & 0.56   & 1.24  &\\
CZTiS    &   KS-I    &   5.372    &  0.993     & 72.4  & 0.64   & 1.32  &\\
CZTiS    &   ST-I    &   5.403    &  1.011     & 40.8  & 0.49   & 1.14  &\\
\hline
\hline
\end{tabular}
\end{table}

The calculated band gap of CZTiSe is smaller than CZTiS in the same crystal structure, which is consistent with the common expectation that the selenides have smaller band gaps than the corresponding sulfides. From the EVU calculations, estimated energy gaps E$_g$ are 1.42 eV for KS CZTiS, 1.24 eV for ST CZTiS, 1.24 eV for KS CZTiSe, and 1.06 eV for ST CZTiSe, respectively. The KS ones have the largest fundamental band gaps among the different cation configurations. The gap values are found to increase more or less linearly upon increasing U value. For example, if Ti has U$_{eff}$=3 eV, the band gap will decrease about 0.05 eV. Thus, the calculated band gaps have an error bar with order of magnitude 0.1 eV. It should be important to compare the theoretical results with experimental data in the future. Nevertheless, as these U are typical values used in the literature\cite{czsse-gapldau}, it would be reasonable to estimate the gap of CZTiS and CZTiSe. Furthermore, the effect of different cationic occupation on the band gap is indicated by comparing four different crystal structures KS, ST, KS-I and ST-I. Within EVU, we find the band gap of KS-I CZTiS is 1.32 eV, whereas that of ST-I CZTiS is 1.14 eV. Likewise, substitution positions induce lattice constant to change and the lattice constant in turn affects energy band gap closely. As there is no available data of lattice and gaps experimentally, it is highly expected to identify structural properties of those materials, and further control disordering and atomic concentration to adjust band gaps.

\begin{figure}
\includegraphics[width=7cm,angle=0]{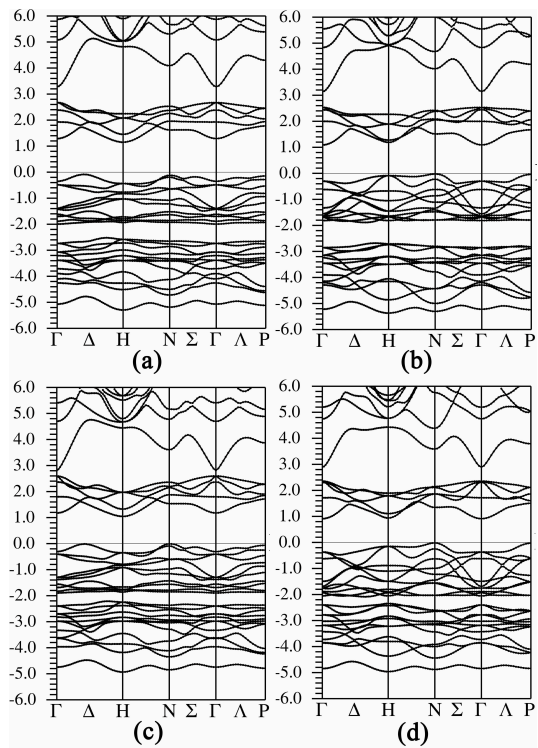}
\caption{\label{sum} Band structure of the (a) KS CZTiS, (b) ST CZTiS, (c) KS CZTiSe and (d) ST CZTiSe, calculated with the DFT+U method. The Fermi energy is set at zero. The high symmetry lines is along $\Gamma$ (0 0 0) - H (1 0 0) - N ($\frac{1}{2}$ $\frac{1}{2}$ 0) - $\Gamma$ (0 0 0) - P ($\frac{1}{2}$ $\frac{1}{2}$ $\frac{1}{2}$). }
\end{figure}

\begin{figure}
\includegraphics[width=8cm,angle=0]{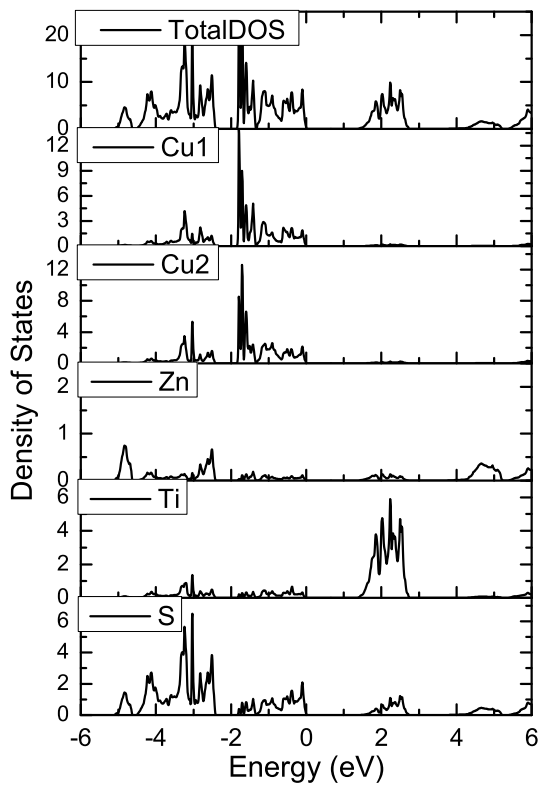}
\caption{\label{sum} The total DOS and partial DOS of KS CZTiS.}
\end{figure}

The band structures of KS CZTiS, ST CZTiS, KS CZTiSe and ST CZTiSe are shown in Fig.2, respectively. CZTiS and CZTiSe show rather similar band structures and all of them have indirect band gaps. The calculations indicate that the highest occupied band are mainly $d_{xy}$ state of Cu while the lowest unoccupied band are maily $d_{3z^2-1}$ state of Ti. The minimum of direct band gap is found on the $\Gamma$ - H line. It is 1.6 eV and 1.4 eV for KS CZTiS and CZTiSe at $\Gamma$, respectively. While the minimum is at H point for ST compounds, it is 1.3 eV and 1.2 eV for CZTiS and CZTiSe, respectively. The different gap behavior should be due to different local coordination of second nearest neighbor cations: In KS lattice, both Cu are surrounded by four tetrahedral coordinated Cu, however one is surrounded by square coordinated Ti and tetrahedral coordinated Zn, whereas the other is surrounded by square coordinated Zn and tetrahedral coordinated Ti. While in ST lattice, every Cu has four square coordinated Cu, but tetrahedral coordinated Zn and Ti. Thus band structures have different dispersion relationships.

Density of states (DOS) of KS CZTiS is shown in Fig.3. It is found that below -4 eV and above 4 eV, there are mainly S 3$p$ states hybridized with $s$ states of Zn, corresponding to bonding and anti-bonding states, respectively. From -4 eV to -2.5 eV, there are bonding states which consist of S 3$p$ states hybridized with 3$d$ states of two Cu, while from -2 eV to -0 eV, there are mainly anti-bonding states which consist of S 3$p$ states hybridized with 3$d$ states of two Cu. The $p$-$d$ bonding and anti-bonding states are separated by a narrow gap about 0.5 eV. Beyond Fermi level, there is a peak of Ti 3$d$ states up to 3 eV. The calculations indicate that the band gap originates from the full occupied Cu $d_{xy}$ valence bands and unoccupied $d_{3z^2-1}$ conducting bands.

As shown in Table I, it indicate that KS and ST CZTiS (or CZTiSe) have comparable energy gaps: E$_g$(KS)-E$_g$(ST)=0.18 eV. The discrepancy of gap size induced by different crystal structure is similar with that of CZTS or CZTSe\cite{czsse-gaphse06,czsse-gapldau}. However, comparing the band gaps of lighter sulfides with the corresponding structure of the selenides, it is found that E$_g$(CZTiS)-E$_g$(CZTiSe) is about 0.2 eV. The values are smaller than that between CZTS and CZTSe, which is about 0.5 eV\cite{czsse-gaphse06,czsse-gapldau}. The reason is that the two series of semiconductors have different gap characters. The direct band gaps of CZTS and CZTSe are derived from the Sn $s$ state and Cu 3$d$ states, in which the position of Sn $s$ band is controlled by anti-bonding between Sn $s$ state and $p$ states of S or Se. Whereas the indirect gaps of CZTiS and CZTiSe consist of 3$d$ states of Ti and Cu, therefore it is much influenced by the $d$-$d$ interaction, controlled more by U term.

\begin{figure}
\includegraphics[width=6cm,angle=270]{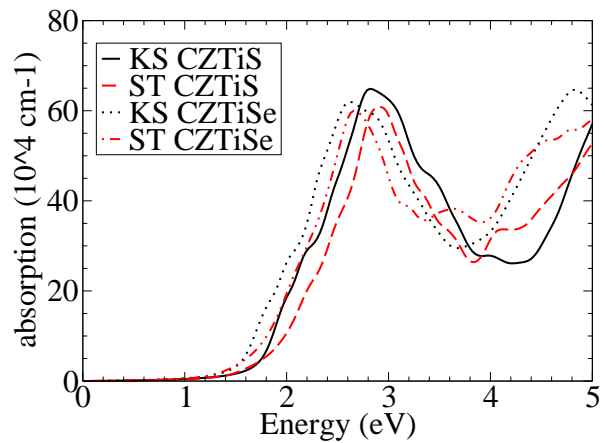}
\caption{\label{sum} (Color online) Absorption coefficients of KS CZTiS (balck solid lines), ST CZTiS (red dash lines), KS CZTiSe (balck dot lines) and KS CZTiSe (red dot-dash line) demonstrating the large band-edge absorption of these materials. The spectra include a 20 meV Lorentzian broadening. ST CZiTS and CZTiSe have similar spectra as the corresponding KS structures.}
\end{figure}

In Fig.4, it shows the absorption coefficient $\alpha(\omega)$ of KS CZTiS, ST CZTiS, KS CZTiSe and ST CZTiSe, respectively, which represents the linear optical response from the valence bands to the lowest conducting bands. All four compounds have comparable absorption, although with different photon energy for the onset to absorption (i.e., the band-gap energy). It is found that KS CZTiS has a large band-edge absorption coefficient. Notably, the calculated values is almost twice larger than that of CZTS and CZTSe\cite{czsse-gapldau}. That imply Ti-compounds might have higher light transformation efficient then Sn-compounds. For higher energies (about 2.5 - 4.0 eV), the absorptions in both CZTiS and CZTiSe are decreasing but it is still similar with CZTS and CZTSe, although the conducting band characters are different. The calculated absorption coefficient is limited by considering direct transitions only, and at high temperatures phonon-assisted transitions will be an additional contribution that will somewhat broaden the spectra. Therefore, it is worthy to extensively investigate CZTiS and CZTiSe to search the possible high-efficient materials, especially in sulfides.

\section{SUMMARY}
Our work demonstrates that the band gaps of CZTiS and CZTiSe originate from the full occupied Cu $3d$ valence band and unoccupied Ti $3d$ conducting band, which is quite different from the well known I2-II-IV-VI4 quaternary semiconductors CZTS and CZTSe. CZTiS and CZTiSe might be candidate materials for solar cell absorber since their band gaps are around 1.2 eV$\sim$1.4 eV, which is comparable with CZTS and CZTSe, although we should note again that the size of gap is depended on the choice of U. As these materials have higher absorption coefficient theoretically, it might be convenient for industrial application not only because Ti is extensively used and cheaper than Sn.

\acknowledgments
This work is supported  by Nature Science Foundation of China (Grant No. 10704024), by Shanghai Rising-Star Program (Grant No. 08QA14026), and by Shanghai Scientific Project
(Grant No. 08JC1408400).

\bibliography{your-bib-file}

\begin{references}

\bibitem{solar1} P. Jackson, D. Hariskos, E. Lotter, S. Paetel, R. Wuerz, R. Menner, W. Wischmann and M. Powalla, Prog. Photovolt: Res. Appl. {\bf 19}, 894 (2011)

\bibitem{czts1} Y. K. Kumar, G. S. Babu, P. U. Bhaskar and V. S. Raja, Sol. Energy
Mater. Sol. Cells {\bf 93}, 1230 (2009).

\bibitem{czts2} L. Wahab, M. El-Den, A. Farrag, S. Fayek and K. Marzouk, J.
Phys. Chem. Solids {\bf 70}, 604 (2009).

\bibitem{czts3}  J.H. Shi, Z. Q. Li, D. W. Zhang, Q. Q. Liu, Z Sun and S. M. Huang, Prog. Photovolt: Res. Appl. {\bf 19}, 160 (2011)

\bibitem{czts4} K. Jimbo, R. Kimura, T. Kamimura, S. Yamada, W. S. Maw, H.
Araki, K. Oishi and H. Katagiri, Thin Solid Films 515, 5997 (2007).


\bibitem{czsse-gaphse06}
S. Y. Chen, X. G. Gong, A. Walsh and S. H. Wei, Appl. Phys. Lett. {\bf 94}, 041903 (2009)

\bibitem{czsse-gapldau} C. Persson, J. Appl. Phys. {\bf 107}, 053710 (2010)

\bibitem{tio2} B. O¡¯Regan and M. Gr$\ddot{a}$tzel, Nature {\bf 353}, 737 (1991)

\bibitem{wien2k}
P. Blaha, K.Schwzrz, G. K. H. Madsen, D. Kvasnicka, and J. Luitz, WIEN2K, An Augmented Plane Wave+Local Orbitals Program for Calculating Crystal Properties (Technical University Wien, Austria, 2001), ISBN3-9501031-1-2.

\bibitem{pbe96} J. P. Perdew, K. Burke and M. Ernzerhof, Phys. Rev. Lett. {\bf 77}, 3865 (1996)

\bibitem{POT_EV} E. Engel and S. H. Vosko, Phys. Rev. B {\bf 47}, 13164 (1993)

\bibitem{EVUapp} C. Persson and A. Zunger, Phys. Rev. B 68, 073205 (2003); Appl. Phys. Lett. 87, 211904 (2005); C. Persson, C. Platzer-Bj$\ddot{o}$kman, J. Malmstr$\ddot{o}$, T. T$\ddot{o}$ndahl, and M. Edoff, Phys. Rev. Lett. 97, 146403 (2006)

\bibitem{enetrend} S. Y. Chen, X. G. Gong, A. Walsh and S. H. Wei, Phys. Rev. B {\bf 79}, 165211 (2009)



\end{references}

\end{document}